\documentclass[prd, amsfonts, twocolumn, nofootinbib, showpacs]{revtex4}
\usepackage{graphicx}
\usepackage{epsfig}
\usepackage{color}
\usepackage{amsmath}
\usepackage{amssymb}
\newcommand{\be}{\begin{equation}}
\newcommand{\ee}{\end{equation}}
\newcommand{\bea}{\begin{eqnarray}}
\newcommand{\eea}{\end{eqnarray}}

\newcommand{\gapp}{\mathrel{\raise.3ex\hbox{$>$}\mkern-14mu
\lower0.6ex\hbox{$\sim$}}}
\newcommand{\lapp}{\mathrel{\raise.3ex\hbox{$<$}\mkern-14mu
\lower0.6ex\hbox{$\sim$}}}
\def\bbox{{\,\lower0.9pt\vbox{\hrule \hbox{\vrule height 0.2 cm
\hskip 0.2 cm \vrule  height 0.2 cm}\hrule}\,}}

\begin{document}
\title{variance of Newtonian constant from local gravitational acceleration measurements}
%\author{}
%\affiliation{ }
\author{De-Chang Dai$^{1,2}$\footnote{communicating author: De-Chang Dai,\\ email: diedachung@gmail.com\label{fnlabel}}}
\affiliation{$^1$ Center for Gravity and Cosmology, School of Physics Science and Technology, Yangzhou University, 180 Siwangting Road, Yangzhou City, Jiangsu Province, P.R. China 225002 }
\affiliation{ $^2$ CERCA/Department of Physics/ISO, Case Western Reserve University, Cleveland OH 44106-7079}

 %%%%%%%%%%%%%%%%%%%%%%%%%%%%%%%%%%%%%%%%%%%%%%%%%%%%%%%

\begin{abstract}
\widetext

We use IGETS absolute gravitational acceleration measurement data to study the gravitational acceleration variance. The relative variance of $\delta g /g$ in 22 years is less than $4\times 10^{-8}$. Since $\delta G /G\lessapprox\delta g /g $, this implies the relative variance of Newtonian constant is less than $3\times 10^{-9}$ based on an sine-like oscillation hypothesis. This limit is at least 4 orders of magnitude better than the existing $G$ measurements. The scattered values of reported $G$ measurements coming from different experiments are most probably coming from systematic errors associated with these experiments and not due to intrinsic time variation of $G$. We also find that $\dot{ G} /G<5.61\times 10^{-10} \text{yr}^{-1}$ based on a linear hypothesis. This is the best terrestrial result so far.   

\end{abstract}

%%%%%%%%%%%%%%%%%%%%%%%%%%%%%%%%%%%%%%%%%%%%%%%%%%

\maketitle

\section{introduction}

The gravitational interaction was first  formulated by Newton in 1687\cite{Newton}. It was the first description of a universal fundamental interaction.  Newton's formula is still valid in most terrestrial experiments, and only small corrections are needed when describing the solar system dynamics and even some phenomena on galactic scales. To study even larger scales or  strong fields and high velocities regimes, Einstein's General Relativity must be used. General relativity has also been subjected to numerous tests and observations. At largest scales, these observations reveal that there are at least two kinds of unknown substances in our universe - dark energy and dark matter.

\begin{figure}
   %\centering
\includegraphics[width=8cm]{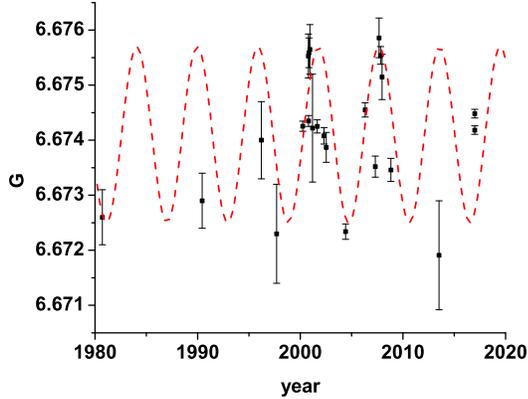}
\caption{ Black squares and errors are the measurement of Newton's constant from 1980 to 2018. The measurement period and errors are from the table II from \cite{Schlamminger:2015hqa}.The two data points in the year 2016 are extracted from \cite{2018Natur.560..582L,Wuhang}. The measurement are taken from 2014 to 2017. We use 2016 as an average of the period. The red dashed curve is $6.6741+0.0016\cos(2\pi t/5.9-0.7)$. The measurement seems to follow a 5.9 years period oscillation. The red curve is plotted to lead the reader's eye, its parameters do not come from the fitting of the data. } 
\label{G-year}
\end{figure}

Gravity is sourced by the matter and in turn affects the spacetime geometry in the universe. The strength of gravity in both Newton's theory and General Relativity is determined by Newton's constant, $G$. The precise value of $G$ is very important for theoretical and practical reasons.  There are more than 200 $G$ measurement experiments after the first value been obtained by Henry Cavendish in 1798\cite{1987Metro..24...44G}. In 2015, the Committee on Data for Science and Technology (CODATA) published a recommended value with a relative standard uncertainty of $4.7\times 10^{-5}$ ( In 2018, the precision achieved $2.2\times 10^{-5}$ according to CODATA.  ). However, the published values are scattered significantly more than the reported uncertainty (fig \ref{G-year})\cite{Speake2014}. This might imply that there are some hidden systematic errors present in most of the experiments. In 2015, Anderson et. al. found that the measured  values of $G$ oscillate with a period $T=5.9\text{yr}$ and amplitude $\approx 1.6 \times 10^{-14}\text{m}^3 \text{kg}^{-1} \text{s}^{-2}$, which was later confirmed by Schlamminger et. al. \cite{Speake2014,Anderson:2015bva,Schlamminger:2015hqa}. The ratio of the offset amplitude, $A\approx 2\times 10^{-4}$  is far larger than the reported random and systematic errors  ($\approx 4.7 \times 10^{-5}$). Anderson et. al. mentioned that the only known phenomenon  with a 5.9 year period is the length of the day, but it does not seem likely that the reason behind it are activities of the Earth's core. Further investigation of this issue  is warranted.

Apart from the technical and scientific issues,  there are two extra factors that might have some implications. First, many experiments were performed by small groups with little or no previous experience on the topic. Second, there are no two identical experiments that have been repeated. Most of the researchers are focusing on new methods, for example in \cite{Armano:2019cac}. The National Institute of standards and Technology, the International Union of Pure and Applied Physics, and international Committee for Weight and Measures invested a significant effort in finding the source of the discrepancies\cite{NIST,IUPAP,CIPM}.  They proposed a potential approach for solving this discrepancy. The method is measuring $G$ with many different methods at the same place. This may eliminate the systematical error induced by the location and experimentalists.  The experimental group in Huazhong University of Science and Technology (HUST) has adopted this idea, performed two independent experiments, and reduced the uncertainty to $1.1\times 10^{-5}$\cite{2018Natur.560..582L,Wuhang}.  This method showed that two independent experiments gave very close results, but they are still outside their own error bars (one is $6.674184(78)\times 10^{-11}\text{m}^3 \text{kg}^{-1} \text{s}^{-2}$ and the other is $6.674484(78)\times 10^{-11}\text{m}^3 \text{kg}^{-1} \text{s}^{-2}$).  More reviews of measurement of $G$ can be found in \cite{Wuhang,Wu:2019pbm,2017RScI...88k1101R} .

There are several constraints on $G$ variance from celestial experiments. The supernova data shows the variance is less than $\dot{G}/G<10^{-10}\text{yr}^{-1}$ in 9 Gyr\cite{Mould:2014iga}. The analysis of transit times in exoplanetary systems gives a constraint \cite{Masuda:2016ggi} $\dot{G}/G<10^{-6}\text{yr}^{-1}$. The best current measurement is $|\dot{G}/G|$ are  $<10^{-13}\text{yr}^{-1}$, which is based on two years data from MRO spacecraft\cite{2011Icar..211..401K}. However, all these measurements are not terrestrial and some of them are model dependent. Therefore it is worth to test how $G$ is changing on Earth. We use the absolute gravitational acceleration provided by International Geodynamics and Earth Tide Service(IGETS)\cite{IGET,Boy1,Boy2}. The measurement stations are mainly located in France, and can be also found in many places around the world. Our analysis indicates that the amplitude must be smaller than $3\times 10^{-9}$, based on an sine-like variance hypothesis, if the oscillating period is less than 20 years. If we adopt a linear variance hypothesis the variance is less than $5.61\times 10^{-10} \text{yr} ^{-1}$. These limits are more stringent than those found by Anderson et. al. This pretty much rules out a $10^{-4}$ variance found by Anderson et. al. We introduce our data and methods in the following.

\section{Data and analysis}

The absolute gravity (AG) time series are from International Geodynamics and Earth Tide Service(IGETS)\cite{IGET,Boy1,Boy2}. The data are measured by the different French instruments, e.g. FG5 206 and 228, operated by Strasbourg and Montpellier teams respectively. It is based on the free falling measurements. The final value is an average of all free falling experiments for a period longer than 12 hours. The diurnal tidal effect is strongly reduced. The locations include France, Africa, Antarctica and Svalbard. 

\begin{figure}
   %\centering
\includegraphics[width=8cm]{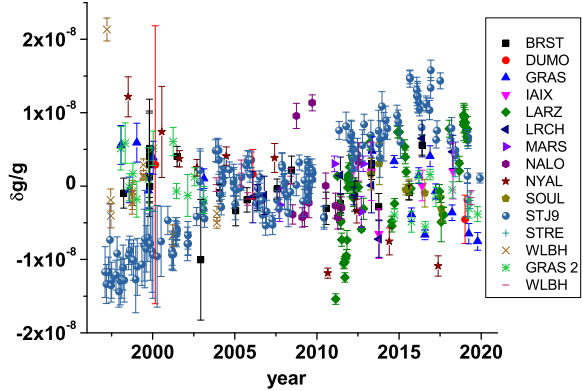}
\caption{  The absolute gravity (g) time series for several sites in France and worldwide, measured by the different French instruments, e.g. FG5 206 and 228 operated by Strasbourg and Montpellier teams respectively, $\frac{\delta g}{\bar{g}}$\cite{IGET}. The observation station are labeled on the right and can be found in \cite{IGET}. The variances is less than $10^{-8}$ in 22 years. This is much smaller than the variance of $G$ measurement. GRAS 2 and WLBH are marked cave in the original data. } 
\label{measurement}
\end{figure}

Data from sixteen observation sites are used in the analysis. Fig. \ref{measurement} shows the variance of gravitational acceleration according to the observation sites, 

\begin{equation}
\frac{\delta g}{\bar{g}} = \frac{ g -\bar{g}}{\bar{g}}.
\end{equation}  

$g$ is the local absolute gravitational acceleration, while $\bar{g}$ is the average of $g$ in the operation period. The data starts from 1997 to 2019. The variance interval is between $-2 \times 10^{-8}$ and $2 \times 10^{-8}$.  This is much smaller than the variance of the measurements of Newton's constant. The variances could be caused by many possible factors, i.e. tides, Earth's polar motion, underground water system, instrumental effects (speed of light, gradient height, vertical transfer), and of course by the fundamental variance of Newton's constant.  

\begin{equation}
\frac{\delta g}{\bar{g}}=\frac{\delta G}{\bar{G}}+\sum_i \frac{\partial_{X_i} g }{\bar{g}}\delta X_i
\end{equation}

$X_i$ represent all the possible factors except for Newton's constant. For example, some values of of the data are increasing with time (for example STJ9) and some are decreasing (for example NYAL), which may be caused by local water systems. This introduces some of the errors in the analysis.  
Since the variance of Newton's constant is just one of the factors of local gravitational acceleration measurement, one expects that its variance must be smaller than variance of the local gravitational acceleration, unless some very fine tuned cancellations are present. This is the reason we can put a very good constraint on the variance of Newton's constant through the local gravitational acceleration measurements. The variance of Newton's constant is estimated from the relation,

\begin{equation}
\label{variance}
\frac{\delta G}{\bar{G}}\lessapprox\frac{\delta g}{\bar{g}}
\end{equation}

We proceed with two hypothesis tests in the following.

\subsection{Period oscillation hypothesis}

According to \cite{Anderson:2015bva,Schlamminger:2015hqa}, Newton's constant appears to follow a 5.9 years oscillation. In order to test this statement, we introduce a test function 

\begin{equation}
f_i(t)=a_i + A_c \cos \omega t +A_s \sin \omega t
\label{cos-fit}
\end{equation}

$i$ is the index to each site. $\omega=2\pi/T$. $T$ is the period. $a_i$ is a constant. Since the average of the gravitational acceleration at different sites can induce different systematic errors, $a_i$ should not be considered as a global constant. Each site has its own value.  $\chi^2$ is calculated according to  

\begin{equation}
\chi^2= \sum_{i,j} \Big(y_i (t_j)-f_i(t_j)\Big)^2.
\end{equation}

$y_i (t_j)$ is the gravitational variance, $\frac{\delta g}{\bar{g}}$, at $i$'th site at time $t_j$. IGETS provides an error of each measurement. In long term these errors can be smaller than the local environmental effect. Therefore, we do not use these error in the analysis.   

\begin{figure}
   %\centering
\includegraphics[width=8cm]{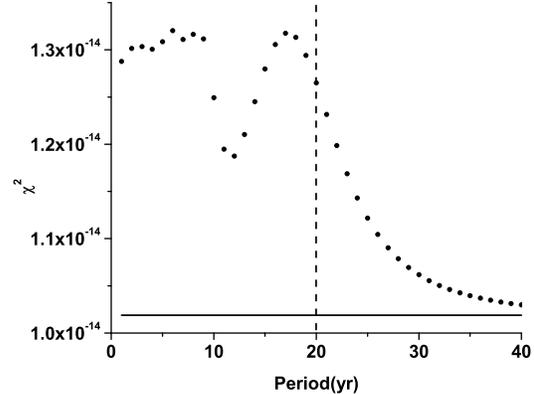}
\caption{ The black points are the $\chi^2$ for fitting function eq. \ref{cos-fit}. The black represents the $\chi^2$ from eq. \ref{linear-fit}. The vertical dashed line marks the period $T=20$. Periods larger than 20 years are not reliable.} 
\label{chi}
\end{figure}

\begin{figure}
   %\centering
\includegraphics[width=8cm]{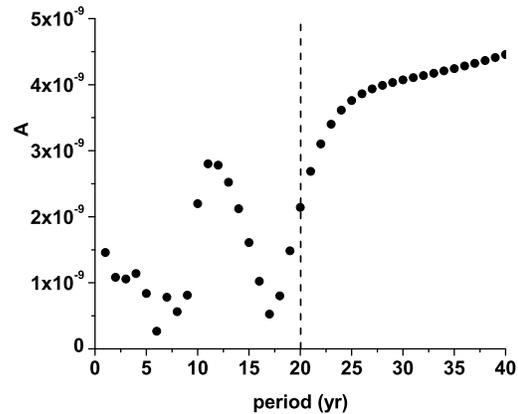}
\caption{  The black points are the amplitude, $A=\sqrt{A_c^2+A_s^2}$, for the fitting function eq. \ref{cos-fit}.  The vertical dashed line marks the period $T=20$.  The amplitudes of the oscillation with period larger than 20 years are not reliable.} 
\label{amplitude}
\end{figure}

Fig. \ref{chi} shows the minimum $\chi^2$ with respect to the period $T$. There is a local minimum at around $T=11 \text{yr}$, but there is nothing special at around $T=5.9\text{yr}$. That means there is no significant signal with this period. Also, this $\chi^2$ is much larger than linear case. The minimum $\chi^2$ decreases quickly for $T$ larger than 20, but it is unreliable, because $T$ is larger than the observation period. So the period hypothesis is not as good as a linear growing hypothesis. Fig. \ref{amplitude} shows the amplitude, $A=\sqrt{A_c^2+A_s^2}$, of the minimum $\chi^2$ with respect to the period $T$. $A$ is about a few $10^{-9}$, which is much smaller than $10^{-4}$ scattering in Newton's constant measurement. Therefore, the scattering of Newton's constant measurements can only be induced by a systematical error.

\subsection{Linear change hypothesis}

Since linear fitting function gives a better result than the period oscillation model, it is instructive to put a constraint on how Newton's constant grows with time. The linear function is chosen to be 

\begin{equation}
fl_i(t)=a_i + b (t-2000)
\label{linear-fit}
\end{equation}

Again $a_i$ is the relative deviation of the relative gravitational acceleration at each site at $t=2000$. The value $2000$ is chosen for convenience, and there is no particular reason to choose it as the beginning year. Again $\chi^2$ is calculated from 

\begin{equation}
\chi^2= \sum_{i,j} \Big(y_i(t_j)-fl_i(t_j)\Big)^2
\end{equation}

The error is estimated from 

\begin{equation}
\sigma^2 = \frac{\min(\chi^2)}{N}
\end{equation}
here, $N$ is the total number of  data points. We calculate the likelihood according to 

\begin{equation}
P=\exp(\frac{-\chi^2}{2\sigma^2}).
\end{equation}

\begin{figure}
   %\centering
\includegraphics[width=8cm]{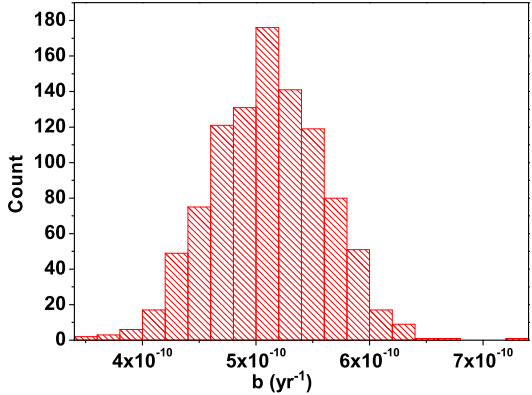}
\caption{ The distribution of $b$. The value is $5.12 \times 10^{-10} \pm 4.9\times 10^{-11} \text{yr}^{-1}$.} 
\label{linear}
\end{figure}

$a_i $ and $b$ are chosen to be flat distributions. 
Markov Chain Monte Carlo are applied according to the probability function and the priors. After stacking all possible values of $a_i$, the distribution of $b$ is shown in fig. \ref{linear}. $b=5.12 \times 10^{-10} \pm 4.9\times 10^{-11} \text{yr}^{-1}$

 From eq. \ref{variance}, $\delta g/g$ can be treated as an upper bound of $\delta G/G$. Therefore the variance of Newton's constant can be estimated as 

\begin{equation}
\frac{\dot{ G}}{G} \lessapprox b+\sigma_b =5.61\times 10^{-10} \text{yr} ^{-1}
\end{equation} 

The dot stands for $\frac{d}{dt}$. This is the best estimate in the terrestrial measurements so far.

\section{Discussion}

We used IGETS absolute gravitational acceleration measurements to study the gravitational acceleration variance based on two hypotheses. First one is the periodic sine-like oscillation hypothesis. If the period is less than 20 yr, the relative variance amplitude of the oscillation  cannot be bigger than $3\times 10^{-9}$. This limit is much stronger than  Anderson et. al.'s finding\cite{Anderson:2015bva}. 
    We also considered a linear variance hypothesis. Under such hypothesis, $\dot{ g} /g=5.12 \times 10^{-10} \pm 4.9\times 10^{-11} \text{yr}^{-1}$. This implies $\dot{ G} /G<5.61\times 10^{-10} \text{yr} ^{-1}$, which is the best terrestrial result so far. This result can further be improved if the equipment is better calibrated, and tidal, ocean loading and other systematic effects are removed. Since we do not have the detailed information of the experimental setup and local environment, we will leave this to future study.  

Our results imply that Newton's constant does not vary significantly in time, and the the scatter in different previous measurements most likely come from systematic errors associated with different methods. 

\begin{acknowledgments}
 D.C Dai is supported by the National Natural Science Foundation of China  (Grant No. 11775140).
\end{acknowledgments}

\end{document}